\documentclass[useAMS,usenatbib]{mn2e}
\usepackage{epsfig,subfigure}
\usepackage{latexsym,amssymb}

\newcommand{\boldsymbol}[1]{\mbox{\boldmath{$#1$}}}
\newcommand{\Msun}{\ensuremath{{\rm M}_\odot}}
\newcommand{\vecr}{\ensuremath{\vec{\boldsymbol r}}}
\newcommand{\vecrp}{\ensuremath{\vec{\boldsymbol r'}}}
\newcommand{\R}{\ensuremath{\mathcal{R}}}
\newcommand{\ev}{\ensuremath{\vec{\boldsymbol{e}}}}
\newcommand{\hv}{\ensuremath{\vec{\boldsymbol{h}}}}
\newcommand{\qhat}{\ensuremath{\hat{\boldsymbol q}}}
\newcommand{\hhat}{\ensuremath{\hat{\boldsymbol h}}}
\newcommand{\ehat}{\ensuremath{\hat{\boldsymbol e}}}
\newcommand{\xhat}{\ensuremath{\hat{\boldsymbol x}}}
\newcommand{\yhat}{\ensuremath{\hat{\boldsymbol y}}}

\begin{document}

\title[The Kozai Mechanism in the GC]{The Effectiveness of
  the Kozai Mechanism in the Galactic Centre} 
\author[Chang]{Philip Chang$^{1}$\thanks{E-mail:
    pchang@astro.berkeley.edu}\\ $^{1}$ Department of Astronomy,
  601 Campbell Hall, University of California, Berkeley, CA
  94720}


\maketitle

\begin{abstract}
I examine the effectiveness of Kozai oscillations in the centres of
galaxies and in particular the Galactic centre using standard
techniques from celestial mechanics.  In particular, I study the
effects of a stellar bulge potential and general relativity on Kozai
oscillations, which are induced by stellar discs.  
\cite{Lockmann2008b} recently suggested that Kozai oscillations induced by the two young
massive stellar discs in the Galactic centre drives the orbits of the
young stars to large eccentricity ($e\approx 1$).  If some of these
young eccentric stars are in binaries, they would
be disrupted near pericentre, leaving one star in a tight orbit around
the central SMBH and producing the S-star population. I find that the {\it spherical} stellar bulge suppresses
Kozai oscillations, when its
enclosed mass inside of a test body is of order the mass in the
stellar disc(s).  Since the stellar bulge in the Galactic centre is much 
larger than the stellar discs, Kozai oscillations {\it due to the
  stellar discs} are likely suppressed.  Whether Kozai oscillations are
induced from other nonspherical components to the potential (for
instance, a flattened stellar bulge) is yet to be determined.
\end{abstract}


\section{Introduction}\label{sec:intro}

Within 0.04 pc from Sgr A$^*$, a swarm of young stars, known as the
S-stars\citep{Eckart1997} or S0-stars \citep{Ghez2005}, orbit the
central supermassive black hole (SMBH) in highly eccentric Keplerian
orbits with random inclination, i.e. an isotropic distribution.  These
S-stars are typically of type B2 with a mass of $\sim 10\Msun$ and a
main sequence lifetime $\sim 20$ Myrs \citep{Alexander2005}.  Outside
of these S-stars (0.04 pc to 0.4 pc), another population of young,
massive stars, which consist of $\sim 6\pm 1$ Myr O- and early B-type
stars, have a very different distribution.  Rather than being
isotropic, these stars are organized into one \citep{Lu2006} or two
disc(s) \citep{Paumard2006}.  

The exact number of discs remains
controversial. \cite{Paumard2006} find that these young stars are arranged in two
stellar discs, which are 
inclined at $\approx 115$ degrees relative to one another
\citep{Paumard2006,Levin2003,Genzel2003}.  The stars in one disc
(which is the more massive disc) have roughly circular orbits with the
typical eccentricity $<0.4$, while the stars in the other less massive
disc are very eccentric ($e>0.6$).  On the other hand \cite{Lu2006}
find only one disc of stars, the more massive disc which
\cite{Paumard2006} identifies, and a ``halo'' population of young
stars in random, highly inclined, and eccentric orbits.

The origin of the young stars in the central parsec of the Galaxy
remains an unsolved problem \citep{Alexander2005}.  For the young
stars between 0.04 and 0.4 pc, their arrangement in a disc suggests
that they may have formed in-situ from the condensation of a gas disc
\citep*{Levin2003,Nayakshin2005b,Nayakshin2007,Levin2007}.  Such star
formation would be expected in accretion discs at large radii due to
the fragmentation from self-gravity
\citep*{Paczynski1978,Kolykhalov1980,Goodman2003,Thompson2005,Chang2008}.  The
more massive, more securely identified disc seem to favor this in-situ
model as the stellar orbits are circular and dynamically cold
\citep{Paumard2006,Lu2006}.  Alternatively, these stars may have
formed in a large cluster at large radii, which spiraled inward due to
dynamical friction \citep{Gerhard2001,Hansen2003,Berukoff2006}.

The problem of the S-stars' formation is even more daunting
than that of the young, massive stars between 0.04 and 0.4 pc
\citep[for a review see][]{Alexander2005}.  For the S-stars, the tidal
field from the SMBH and their current random orientation argues
against both an in-situ disc formation scenario and a sinking cluster
\citep{Ghez2005}. The S stars may have formed via a different channel
than the young, massive stars.  \cite{Levin2007} has argued that type
1/2 migration to small radii \citep{Goldreich1978}, followed by
resonant relaxation \citep{Rauch1996} is a viable formation
channel. On the other hand, \cite*{Perets2007} argues that the large
population of massive perturbers near the Galactic centre (GC)
increases the relaxation rate to such an extent that these S-stars may
have been formed when binaries, which are scattered into large
eccentricities, are disrupted near pericentre \citep{Hills1988,Gould2003}.

Recently, a rather elegant scenario has been proposed by
\cite*[][hereafter LBK]{Lockmann2008b} for the origin of the S stars
and their link to the young, massive stars which surround them. LBK
showed that the S-stars could be formed from tidal disruption of
binaries \citep{Hills1988,Gould2003} which are driven to large eccentricity from
Kozai oscillations \citep{Kozai1962,Lidov1962} induced from the outer
two young massive stellar discs. Earlier work by Subr, Karas and 
collaborators \citep{Subr2004,Subr2005,Karas2007} studied the case for a 
single disc (with gasdynamic dissipative effects for the case of a fossil 
gas disc). This is indeed a very
attractive proposal as it naturally would explain their isotropic
distribution, large eccentricity, and apparent youth.  Using an N-body
calculation with relativistic correction of up to 2.5 post-Newtonian
orders \citep{Lockmann2008a}, they showed that stars (or binaries)
from the two discs can achieve eccentricities as large as $e\approx
0.999$.  Relativistic corrections, which typically, can damp large
eccentricities in the Kozai mechanism \citep*{Holman1997,Fabrycky2007}
appear not to be important (LBK).  Binaries in these eccentric orbits
that are disrupted near pericentre will result in one of the stars in
a tight orbit around the SMBH.  The reduction of semimajor axis from
such a binary fission event would be of order a factor of ten
\citep{Gould2003} and would be a viable mechanism for the formation of
the S-stars.

The beauty of LBK's scenario is that it ties in observed properties of
the S-star distribution, namely their isotropy, eccentricity, and
small semimajor axis with the observed properties of the young stellar
discs, namely their high inclination of $115$ degrees relative to each
other. However, LBK's result is surprising as the Kozai mechanism 
is fairly delicate and can be greatly suppressed if additional perturbations 
to the gravitational potential such as a stellar cusp are included.
Thus, I am motivated to examine of the basic physics of
this scenario and ask the question: to what degree might Kozai
oscillations be important in the centres of galaxies.  In this paper,
I study the nature of the Kozai mechanism central this
scenario.  I present my basic model and the basic equations for the
Keplerian orbital parameters in \S\ref{sec:secular}.  In
\S\ref{sec:GC}, I apply my basic model to the Galactic Centre.  I
numerically compute the evolution of a star subject to the perturbed
potential from a single disc, a single disc with a stellar bulge, two
discs, and two discs with a bulge.  I show that including the
spherical stellar distribution of sufficient mass suppresses the Kozai
mechanism. I then argue that the spherical stellar distribution in
Galactic centre is more than adequate of this purpose.  I discuss some
of these implications and conclude in \S\ref{sec:conclusions}.

\section{Secular Evolution}\label{sec:secular}

I now describe the model problem central to this study.  Two massive
stellar discs with mass $M_1$ and $M_2$ orbit a central SMBH with
mass $M_0 \gg M_1, M_2$ at an inclination relative to one another of
$i_{0}$.  The surface density of these two stellar discs,
$\Sigma_{1,2}$, is assumed to be a power law with index $-p$, i.e.,
$\Sigma_{1,2} = \Sigma_{1,2,0} (r/r_0)^{-p}$, where $r$ is the radial
coordinate, $r_0$ is a reference radius, and $\Sigma_{1,2,0}$ is the
normalization for disc 1 and 2 respectively.  Without loss of
generality, I orient my axis such that the reference plane is in the
plane of the $M_1$ disc and I presume that $M_1 \ge M_2$.  Also
surrounding the SMBH is a spherical stellar power law distribution
with index $-q$, $n_* = n_0 (r/r_0)^{-q}$.  A test body orbits the
SMBH in a near-Keplerian orbit with semimajor axis, $a$, inclination,
$i$, and eccentricity, $e$.

I first consider the case of a perturbing mass, $\delta m$, in a
circular orbit of radius, $r$, around the central mass and its effect
on the test body. I will
work exclusively in the secular approximation.  This problem is well
studied by many authors \citep[see for
  instance][]{Innanen1997,Kiseleva1998,Ford2000}.  In particular,
\cite{Eggleton1998} \citep[see also][]{Eggleton2001,Fabrycky2007} has
introduced a nice formalism which I find very flexible, powerful, and
especially useful for studying multiple perturbing bodies which are at
large inclinations relative to one another.  Using the notation of
\cite{Eggleton2001} and \cite{Fabrycky2007} and dropping dissipative
and tidal terms, the governing equations are:
\begin{eqnarray}
  \frac 1 e \frac {d\ev}{dt} &=& \left(Z_{\rm GR} + Z_{\rm *}\right)\qhat \nonumber\\
&&- \left(1 - e^2\right)\left[5 S_{eq}\ehat - \left(4S_{ee} - S_{qq}\right)\qhat + S_{qh}\hhat\right], \label{eq:dotev}\\
  \frac 1 h \frac {d\hv}{dt} &=& \left(1-e^2\right)S_{qh}\ehat - \left(4e^2 + 1\right)S_{eh}\qhat + 5e^2S_{eq}\hhat,\label{eq:dothv}
\end{eqnarray}
where $\ev$ is the Laplace-Runge-Lenz vector, whose magnitude is the
eccentricity, $e$, $\hv$ is the reduced orbital angular momentum
vector, $\hhat$ and $\ehat$ are the normalized vectors in the
direction of $\hv$ and $\ev$ respectively, and $\qhat = \hhat \times
\ehat$ is the normal vector that completes the triad.
$(\hhat,\ehat,\qhat)$.  The disturbing tensor\footnote{I am
  unaware of a name for this tensor, so for the sake of nomenclature, I have chosen to call this the disturbing tensor.} is $S_{xy} = C\left[\delta_{xy} - 3(\hhat'\cdot\xhat)(
  \hhat'\cdot \yhat)\right]$, where $\hhat'$ is the normalized angular
momentum vectors of the perturbing mass and the constant, $C$, is
\begin{equation}\label{eq:constant}
C = n(a)\frac {\delta m} {M_0} \left(\frac{a}{r}\right)^3\left(1-e^2\right)^{-1/2},
\end{equation}
where $n(a) = \sqrt{G M_0/a^3}$ is the mean motion.  Finally $Z_{\rm
  GR}$ and $Z_{\rm *}$ represents the apsidal motion from the effects
of general relativity and the stellar cusp.  The precession term due
to general relativity is \citep{Eggleton2001,Fabrycky2007}:
\begin{equation}
Z_{\rm GR} = \frac 3 2 \frac {r_g}{a} \frac n {1-e^2}, 
\end{equation}
where $r_g = 2GM_0/c^2$ is the gravitational radius. The precession
due to the stellar bulge potential is \citep[][see their eq.(14) and
  (15) and appendix A]{Ivanov2005}
\begin{equation}\label{eq:aps_prec}
Z_* = -\kappa n \frac {M_*(a)}{M_0},
\end{equation}
where $M_*(a) = 4\pi\int_0^a m_* n(r) r^2dr$ is the mass of the bulge
stars inside of sphere whose radius is equal to the test body's semimajor axis, $a$, 
\begin{equation}
\kappa = \frac{\Gamma(5/2-q)}{\sqrt{\pi}\Gamma(3-q)}
\end{equation} 
is a constant of order unity, and $\Gamma$ is the gamma function.

I now calculate the effect of a disc.  Taking $\delta m \rightarrow dm
= 2\pi\Sigma r dr$, I find that the effect of a disc is a modification
of equation (\ref{eq:constant}) to be
\begin{equation}
C_{\rm d} = n(a) M_0^{-1} a^3\left(1-e^2\right)^{-1/2}2\pi\int_{r_{\rm in}}^{r_{\rm out}}dr \Sigma r^{-2} dr,
\end{equation}
where $C_{\rm d}$ is the constant associated with the disc, $r_{\rm
  in}$ is the disc inner radius, and $r_{\rm out}$ is the outer radius
of the disc.  For $r_{\rm out} \gg r_{\rm in}$, this gives
\begin{equation}
C_{\rm d} = n(a) \frac {M_{\rm eff}}{M_0} \left(\frac a {r_{\rm in}}\right)^3\left(1-e^2\right)^{-1/2}
\end{equation}
where $M_{\rm eff} = 2/(1+p)\pi\Sigma_{1,0}r_0^2$ is the effective
mass of the disc, and $M_{\rm d} = 2\pi\int_{r_{\rm in}}^{r_{\rm out}}
dr r \Sigma$ is the mass of the disc.  Note that the effect of a
massive extended stellar disc with $r_{\rm out} \gg r_{\rm in}$ can be
reduced to a single ring with mass $M_{\rm eff}$ at a radius of
$r_{\rm 1,2 in}$ for $p > -1$. For $p = -1$, the contribution is
logarithmic in radius and for $p<-1$, the outer radius is more
relevant.  

The use of the governing equations (\ref{eq:dotev}) and
(\ref{eq:dothv}) is especially advantageous for studying the secular
effects of multiple perturbing rings.  Like the Lagrange-Laplace
planetary equations \citep{Brouwer,Murray}, the governing equations
(\ref{eq:dotev}) and (\ref{eq:dothv}) are linear.  Hence, the effect
of additional perturbing masses is just a matter of adding their
associated disturbing function (in the case of the Lagrange-Laplace
planetary equations) or the perturbing potential, $S_{xy}$, in the
case of equations (\ref{eq:dotev}) and (\ref{eq:dothv}).  The
calculation of the {\it inclined} disturbing function in the case of
the secularly averaged Laplace-Langrange's planetary equations
\citep{Brouwer,Innanen1997,Kiseleva1998,Murray,Ivanov2005} is rather
involved and I have found it to be numerically difficult to
solve. By contrast including additional masses is trivial for the
governing equations (\ref{eq:dotev}) and (\ref{eq:dothv}). For a
number of perturbing masses, $\delta m_i$, the disturbing tensor,
$S_{xy}$, is a sum over all disturbing tensors from all the masses, or
\begin{equation}
S_{xy}= \sum_i C_i\left[\delta_{xy}
  - 3(\hhat'_i\cdot\xhat)( \hhat'_i\cdot \yhat)\right],
\end{equation}
where $\hhat'_i$ is the normalized angular momentum vectors of the
perturbing mass $\delta m_i$.  The associated constant, $C_i$, is
\begin{equation}
C_i = n(a)\frac {\delta m_i} {M_0} \left(\frac{a}{r_i}\right)^3\left(1-e^2\right)^{-1/2},
\end{equation}
where $r_i$ is the radial position of the perturbing mass.  

\section{Application to the Galactic Centre}\label{sec:GC}

In the case of the Galactic centre, \citep{Paumard2006} find that the
stellar distribution follows $p\approx 2$, $r_{\rm 1,2,in}=0.1$ pc,
and $r_{\rm 1,2,out} = 0.5$ pc for the two disc, which are oriented at
$115\pm 7$ degrees with respect to one another. The upper limit for the
mass of the clockwise disc is $M_{\rm d,1} < 10^4\Msun$, and the upper
limit for the mass of the counterclockwise disc is $M_{\rm d,2} <
5\times 10^3\Msun$ \citep{Paumard2006}, which is consistent with
\cite{Nayakshin2006}'s dynamical limits of $\lesssim 10^4\Msun$ for
the masses of two discs.  For a top-heavy initial mass function (IMF),
\cite{Paumard2006} derives a lower limit for the stellar mass of the
two discs to be $3500\Msun$ and $1400\Msun$.  Hence for a range in
disc masses between $M_{\rm d,1} = 3500-10^4\Msun$ and $M_{\rm
  d,2}=1400-5000\Msun$, the effective disc masses are $M_{\rm 1,eff}
\approx 0.6-2\times 10^3\Msun$ and $M_{\rm 2,eff} \approx 0.3-1\times
10^3\Msun$.

\begin{figure}
\begin{center}\includegraphics[width=.45\textwidth]{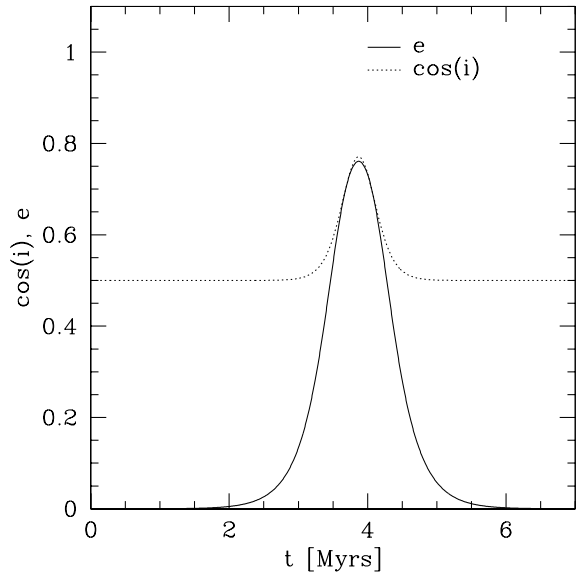}\end{center}
\caption{Simple illustration of the Kozai oscillation for one disc of
  mass, $M_{\rm eff} = 2\times 10^3\Msun$ for a test particle (star) with an
  initial inclination of 60 degrees relative to the plane of the disc.
  The test particle's eccentricity and inclination are shown as a
  solid and dotted lines respectively.  The precession from general
  relativity and the stellar bulge potential are ignored in this
  case.}
\label{fig:simple_case}
\end{figure}

I numerically calculate the evolution of equation (\ref{eq:dotev}) and
(\ref{eq:dothv}), using a standard Runge-Kutta algorithm
\citep{Press1992}.  As an illustration, I plot the eccentricity (solid
line) and inclination (dotted line) for a test particle that is in
orbit around a SMBH and is initially inclined relative at 60 degrees
to a single massive disc with $M_{\rm eff} = 2000\Msun$ in Figure
\ref{fig:simple_case}. For this case, I have taken $Z_{\rm GR} = 0$
and $Z_* = 0$.  Note the characteristic behavior of the Kozai
oscillation, where the inclination and eccentricity vary in phase and
the initial inclination determines the amplitude of
the oscillation due to the conservation of the Kozai integral or
z-component of the angular momentum vector, $L_{\rm z} =
(1-e^2)\cos^2i$ \citep{Kozai1962}.

The axissymmetry, but not spherical symmetry, of the potential
conserves $L_{\rm z}$, but not the total angular momentum vector, $L$.
Thus, an initially inclined low eccentricity orbit achieves very high
eccentricities. For the case of the region
around a SMBH, \cite{Subr2005} and \cite{Karas2007} studied the effect
of fossil gas disc on the orbits of a nuclear star cluster.  In the
case of the a single disc, \cite{Subr2005} and \cite{Karas2007} showed
that there are two kinds of orbits: orbits which librate around
$\omega=\pi/2$ and $3\pi/2$ and orbits which span over $\omega =
[0,2\pi]$, where $\omega$ is the argument of pericentre.  In the particular case of a
fossil gas disc, dissipative interactions from star-crossings of the
fossil gas disc dissipates energy, resulting in a slow decay of the
semi-major axis \citep{Karas2007}.

I now study the effect of Kozai oscillations including the effects of general
relativity, $Z_{\rm GR}$ and the second disc. In agreement with the
LBK's claim, I do not find the effect of relativity to be significant
for eccentricities up to $0.999$.  

\begin{figure}
\begin{center}\includegraphics[width=.45\textwidth]{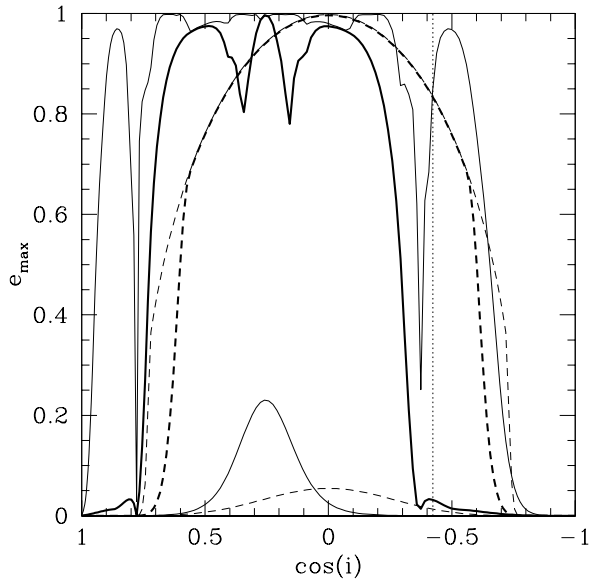}\end{center}
\caption{Maximum eccentricity as a function of initial inclination for
  one disc (dashed lines) with $M_{\rm 1,eff} = 0.6$ (lower dashed
  line), $1.2$ (middle thick dashed line), $2\times 10^3\Msun$ (upper
  dashed line) and two discs with $M_{\rm 1,eff} = 0.6$ (lower solid
  line), $1.2$ (middle thick solid line), $2\times 10^3\Msun$ (upper
  solid line) with respective $M_{\rm 2,eff} = 0.3, 0.6, 1 \times
  10^3\Msun$).  As a function of phase space, note that the regions
  available for large eccentricities ($e\gtrsim 0.95$) is much larger
  for two discs as opposed to one disc.  The vertical dotted line
  indicates the position of the second disc relative to the first for
  the parameters of the two discs in the Galactic centre, i.e., an
  inclination of 115 degrees.}
\label{fig:max_ecc_inclination}
\end{figure}

I now include the effect of the second disc, which is less massive at
$\approx 1.4-5\times 10^3\Msun$ than the first disc ($3.5-10\times
10^3\Msun$).  The observed inclination is fairly narrow $115 \pm 7$
degrees, so I choose to fix the inclination at $115$ degrees.  In
Figure \ref{fig:max_ecc_inclination}, I show the maximum eccentricity,
$e_{\rm max}$, reached as a function of initial $\cos i$ relative to
the reference plane (which is the plane of the disc for the one disc
case and the plane of the more massive disc in the two disc case).
For two discs, $e_{\rm max}$ as a function of $\cos i$ is considerably
more complex than the case for one disc.  Note that for low mass discs
(either for one disc or two disc case), there is insufficient time
($7$ Myrs) for stars to reach high eccentricity.  For two discs, the range
of initial inclinations that generate large eccentricities is larger
compared to the single disc. Also note, that large eccentricities can
be reached for two discs in the neighborhood around the inclination of
the second disc. The addition of a second disc increases the
available inclinations over which test bodies, i.e., stars, can reach
large eccentricities.  Note also that even in the most optimistic
scenario, stars do not oscillate to high eccentricity for $\cos i
\approx \pm 1$.  Stars that are close the to more massive stellar disc
do not reach large inclination in spite of the second disc.  Hence,
stars and binaries that are driven to large eccentricities must come
from the less massive disc, which supports LBK's result that stars in
the CCW disc (the less massive disc) are more likely to be driven to
large eccentricity.

The axissymmetry of a single disc conserves $L_{\rm z}$.  As a result,
a polar plot of $e$-$\omega$ \citep[see for
  instance][]{Subr2005,Karas2007} of an orbit generates a closed
curve.  On the other hand, when a second disc is included,
axissymmetry is broken and an orbit no longer generates close curves
in a polar $e$-$\omega$ plot.

\begin{figure}
\begin{center}\includegraphics[width=.45\textwidth]{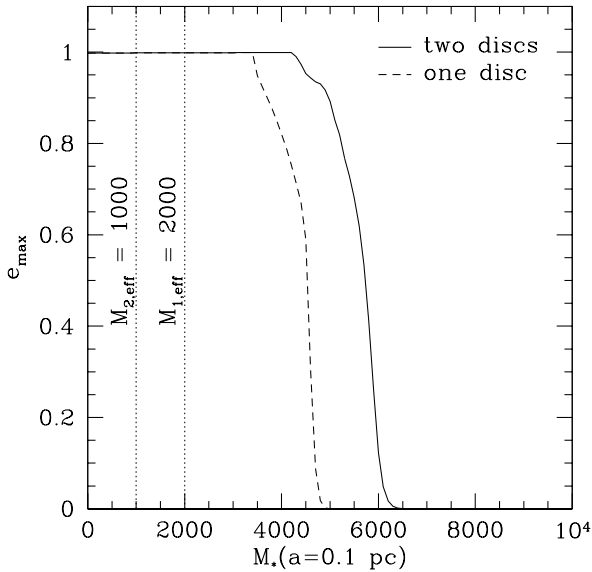}\end{center}
\caption{Maximum eccentricity as a function of stellar mass enclosed
  for $a=0.1\,{\rm pc}$ for one disc (dashed line, $M_{\rm 1,eff} =
  2\times 10^3\Msun$) and two discs (solid line, $M_{\rm 1,eff} =
  2\times 10^3\Msun$ and $M_{\rm 2,eff} = 10^3\Msun$). Also shown is
  $M_{\rm 1,eff}$ and $M_{\rm 2,eff}$ for the comparison of scales. For
  sufficiently large stellar bulges ($M_*(a=0.1\,{\rm pc}) \gtrsim
  3500\Msun$ for one disc and $M_*(a=0.1\,{\rm pc}) \gtrsim 4500\Msun$
  for two discs), large eccentricities cannot be achieved and the
  Kozai mechanism is suppressed.}
\label{fig:max_ecc_simple}
\end{figure}

When the stellar bulge potential is included, $Z_*$, the dynamics
changes completely.  For the bulge potential I take $q=1.4$ (which
gives $\kappa \approx 0.6$ for eq.[\ref{eq:aps_prec}]) and $n_0 =
\rho_0/\Msun$ is the number density of stars, where $\rho_0$ is the
mass density of stars \citep{Genzel2003,Yu2007}.  I plot its effect as
a function of $M_*(a=0.1\,{\rm pc})$ in Figure
\ref{fig:max_ecc_simple} for the most optimistic scenario $M_{\rm
  d,1}=10^4\Msun$ and $M_{\rm d,2}=5000\Msun$.  As this figure shows,
sufficiently large stellar bulges ($M_*(a=0.1\,{\rm pc}) \gtrsim
3500\Msun$ for one disc and ($M_*(a=0.1\,{\rm pc}) \gtrsim 4500\Msun$
for two discs) suppresses the Kozai mechanism.



The mass found that is needed to suppress Kozai oscillations agrees
well with simple analytic arguments.  The Kozai mechanism is known to
be suppressed by non-Keplerian contributions to the potential such as
those introduced by general relativity or additional masses in the
system \citep[see the case for multiple planets,][]{Tremaine2004}.
Typically if the period for apsidal precession is of order or shorter
than the period of the Kozai oscillations, then the Kozai oscillations
are suppressed \citep{Tremaine2004,Fabrycky2007}.  The period of the
Kozai oscillation is of order (from studying eq.[\ref{eq:constant}]):
\begin{equation}
\tau_{\rm K}^{-1} \sim \zeta_{1,2} \sim n \left(\frac a {r_{\rm
    1,2,in}}\right)^{3} \left(\frac {M_{\rm eff}}{M_0}\right).
\end{equation}
whereas the timescale for apsidal precession due to the stellar bulge
is of order (from studying eq.[\ref{eq:aps_prec}])
\begin{equation}
\tau_{*}^{-1} \sim n \frac {M_*(a)}{M_0}.
\end{equation}
Hence, the requirement that the apsidal precession period be longer than
the period for Kozai oscillations imply $M_* \lesssim M_{\rm eff}$,
matching the expectations from the more detailed calculation in Figure
\ref{fig:max_ecc_simple}.

I now argue that the mass of stellar bulge is much larger than what is needed 
to suppress the Kozai mechanism.  From
observations, the measured $M_*(a=0.1\,{\rm pc})$ is $\approx 6\times
10^4\Msun$, using the values for the central stellar density from
\cite{Genzel2003} \citep[see also][]{Schodel2007}.  The observed mass
is over an order of magnitude larger than what is needed to suppress
the Kozai mechanism.  It is significantly larger than the mass of the
two young stellar discs and, therefore, Kozai oscillations induced by the 
two young massive stellar discs are likely suppressed.

The suppression of Kozai oscillations due to a spherical distribution
of stars is well known for the case of a single disc
\citep{Ivanov2005,Karas2007}.  \cite{Ivanov2005} argues that in the
case of a single ring, Kozai oscillations are suppressed for
initially low eccentricity orbits for a sufficiently massive spherical stellar
distribution.  Similarly, \cite{Karas2007}
also showed that a spherical stellar distribution whose mass is
comparable to the perturbing disc mass will suppress Kozai
oscillations for initially circular orbits (see the third panel of
their Figure 2).  The results of this paper are in broad agreement
with these previous results and point out that inclusion of additional
rings, which breaks axial symmetry, do not change the basic results
for small {\it initial} eccentricities.  

\section{Discussion and Conclusion}\label{sec:conclusions}

I find that the apsidal precession induced by the stellar bulge (using a
realistic estimate for its mass) greatly reduces the impact of the 
Kozai mechanism in the GC.  Hence, the Kozai oscillations
central to LBK's elegant mechanism is likely not due to the stellar discs.
The calculation in LBK is more realistic in that it captures the
dynamics of the system without resorting to the perturbative scheme
used in the present study. However, this study is complementary
because it outlines the regions of parameter space where
secular effects are important.

According to the present work, the young stellar disc is not likely to induce 
Kozai oscillations, but other nonspherical 
components to the potential may
able to do so if they are sufficiently strong.  One
possibility is a significantly flatten (of order unity) stellar bulge.
A detailed study of the degree of flattening required to induce Kozai
oscillation is interesting, but it is beyond the scope of this work.
On theoretical grounds, such a significantly flattened bulge may be
unlikely because resonant relaxation \citep{Rauch1996} would
isotropise the bulge stars \citep[see for instance,]{Levin2007} over
their 10 Gyr lifetime.  Observationally, there is no evidence for a
flatten bulge as the velocity distribution of the late-type stars
appear to be consistent with isotropy \citep{Genzel1996}.

The Kozai mechanism may be more applicable in 
other galactic nuclei such as M31, which has a massive stellar disk
(the P1/P2 disk).  In M31, the P1/P2 disc
\citep{Tremaine1995,Peiris2003,Chang2007} is $\sim 10$\% of the mass
of the $1.4\times 10^8\Msun$ SMBH \citep{Bender2005}, whereas the
stellar bulge potential is much smaller, i.e., $M_* <
10^6\Msun$ at 1'' or $\approx 4$ pc \citep{Peiris2003}. \cite{Chang2007} has
suggested that the non-axisymmetric potential of the P1/P2 disc
modifies gas orbits such that they are confined to be inside of 1 pc
around the SMBH.  High inclination stellar orbits may also undergo
Kozai oscillations in this case.  The implications for these stars
undergoing Kozai oscillations in the nucleus of M31 would be an
interesting topic for further study.

\section*{Acknowledgments}

I thank E. Chiang and E. Quataert for encouraging me to study this
issue, for useful discussions, and for detailed readings of this
manuscript. I thank R. Genzel, J. Lu, N. Murray, and G. Van der Ven
for useful discussions and the anonymous reviewer for clarifying
comments and useful suggestions, which greatly improve the 
presentation of this paper.  I am supported by the Miller 
Institute for Basic Research.

\bibliographystyle{mn2e} 
\bibliography{kozai}

\end{document}